\documentclass{epl}
\title{Unitary transformation for the system of a particle in a linear potential}
\author{Dae-Yup Song\inst{1,2}}
\institute{
  \inst{1} Department of Physics, University of Florida, Gainesville, FL 32611, USA\\
  \inst{2} Department of Physics, Sunchon National University, Suncheon
540-742, Korea } 
\begin{document}
\maketitle
\begin{abstract}
A unitary operator which relates the system of a particle in a
linear potential with time-dependent parameters to that of a free
particle, has been given. This operator, closely related to the
one which is responsible for the existence of coherent states for
a harmonic oscillator, is used to find a general wave packet
described by an Airy function. The kernel (propagator) and a
complete set of Hermite-Gaussian type wave functions are also
given.
\end{abstract}
\section{Introduction}
The existence of coherent and squeezed states for a simple
harmonic oscillator \cite{Sch,KS} can be understood from the fact
that there exist unitary transformations which leave the
time-dependent Sch\"{o}dinger equation (formally) invariant under
the transformations. These transformations have been found as the
relations between harmonic oscillators of time-dependent
parameters \cite{Li,Uni}, while the transformations between the
same simple harmonic oscillators can be applied to a stationary
states to give the coherent and squeezed states \cite{Uni}. On the
other hand, Feynman and Hibbs show that the kernel of a general
quadratic system is described by the classical solutions of the
system \cite{FH}. Since the wave function might be derived from
the kernel, this suggests that the unitary transformations are
described by classical solutions, as explicitly  shown in the
quadratic systems \cite{path}.

There has been considerable interest for the system of a particle
in a linear potential (with time-dependent parameters)
\cite{BB,Feng,TZM}. This model has eigenfunctions (wave packets)
described by the Airy function \cite{Landau}. The model and the
Airy wave functions on a half-line have been used to model the
production of high harmonic generation in the laser irradiation of
rare gases\cite{CR}, and the edge electron gas \cite{KM,VJ}. The
model on piecewise domains and the wave functions have been
frequently used to model various physical systems
\cite{piecewise}. The Sch\"{o}dinger equation for a free particle
has also long been interesting in that the equation is formally
identical to the wave equation of a beam of light in the paraxial
approximation \cite{paraxial}.

In this article, we will show that there exists a unitary relation
between the system of a particle in a linear potential and that of
a free particle. Indeed, time-dependent unitary relations have
been known to be useful in analyzing quantum systems. Unitary
transformations have been extensively used in showing that the
Dirac theory goes to the Pauli theory in the non-relativistic
limit \cite{FW}, and a unitary relation between the system of a
charged particle in a purely time-dependent vector potential and
the same system in an accelerated frame without the vector
potential has been given in ref. \cite{KH}. We will find a unitary
transformation which relates the model of a linear potential (with
time-dependent parameters) to a free particle system. The
transformation resembles the one for a quadratic system which
gives coherent states, and could be applied to any wave function
of a free particle to give the wave function of a particle in a
linear potential. This transformation will be explicitly used to
find a general wave packet described by the Airy function, which
clearly shows the origin of the Berry-Balazs solution for the wave
function of a free particle. Based on the kernel and wave
functions of the quadratic system, the kernel (propagator) and a
set of Hermite-Gaussian type wave functions for the linear system
will also be given, and the completeness of the set will be
proved.

\section{The model and a unitary transformation}
The model we will consider is described by the Schr\"{o}dinger
equation
 \begin{equation} i\hbar {\partial \over \partial t}\psi(t,x)=
 -{\hbar^2 \over 2M(t)}{\partial^2 \over
 \partial x^2}\psi(t,x) -xF(t) \psi(t,x),
 \end{equation}
where $M(t)$ and $F(t)$ denote the time-dependent mass and
external force, respectively. Eq. (1) would describe a charged
particle in a linear, scalar potential. The solution $x_p (t)$ for
the classical equation of motion is given as
 \begin{equation}
 x_p = \int_0^t{1\over M(t')} \int_0^{t'} F(t'') dt''dt'+
 C\int_0^t {1 \over M(t')}dt'+D,
 \end{equation}
which satisfies ${d \over dt} (M{dx_p \over dt})=F$, where $C$ and
$D$ are arbitrary real constants. By defining the operators
 \begin{eqnarray}
 O&=&-i\hbar{\partial \over \partial t}+H= -i\hbar{\partial \over \partial t}
 +{p^2 \over 2M}-xF(t), \\
 O_M&=&-i\hbar{\partial \over \partial t}+H_M=-i\hbar{\partial \over \partial t}
 +{p^2 \over 2M},
 \end{eqnarray}
one may find the unitary relation
 \begin{equation}
 U(x_p) O_M U^\dagger(x_p) = O,
 \end{equation}
where
 \begin{equation}
 U(x_p)= \exp\left[{i\over \hbar}[M\dot{x}_p x
 +\xi(x_p)]\right]\exp\left[ - {i \over \hbar} x_p p\right],
 \end{equation}
with the  function of time $\xi$ defined as
 \begin{equation}
 \xi(x_p) = -{1\over 2}\int_0^t\left[ M(t')\dot{x}_p^2(t')\right] dt'.
 \end{equation}
The overdots denote differentiations with respect to the time of
the system.

From this unitary relation, one can find that, if a wave function
$\psi_M$ satisfies the Schr\"{o}dinger equation $O_M\psi_M=0$, the
wave function $U\psi_M$ satisfies the Schr\"{o}dinger equation
$O(U\psi_M)=0$. For a harmonic oscillator system, the center of
probability distribution of a (generalized) coherent state,
obtained from a (stationary) state through a unitary
transformation, moves along the trajectory described by a
classical solution \cite{Sch,KS,Li,Uni}. The unitary relation
between a linear system and the corresponding free particle system
closely resembles the one for (generalized) harmonic oscillators,
in that the probability distribution of the unitarily transformed
wave function moves globally, according to the classical solution,
from the distribution of the original wave function, while the
shapes of the two distributions are same. The shapes could evolve
under the time-evolution, as in the generalized coherent states.
If $F=0$, the unitary relation becomes a relation between the two
physically identical systems, while the relation is still not
identically unity if we choose non-zero $C$ and $D$; In this case,
if $M(t)$ is a constant, the probability distributions move with
the constant speed $C/M$ from each other. As may be clear in the
$F=0$ case, the degrees of freedom of choosing $C$ and $D$ may be
interpreted as a manifestation of Galilean symmetry in quantum
mechanics.
\section{Airy wave packets}For the linear system with a constant mass $m$, a
constant force $f(\equiv {\beta^3 \over 2m})$, and Hamiltonian
$H_m~(={p^2\over 2m} -fx)$, there are eigenfunctions described by
an Airy function:
 \begin{equation}
  \phi={\rm Ai}[-{\beta \over \hbar^{2/3}} (x+e)],
 \end{equation}
with a constant $e$. The Airy function satisfies the equation
 \begin{equation} {d^2 \over dz^2} {\rm Ai} (z) - z {\rm Ai} (z)=0,
 \end{equation}
which gives the energy-eigenvalue relation \cite{Landau}
 \begin{equation}
 H_m\phi(x)=  ({\beta^3 \over 2m}e )\phi(x).
 \end{equation}
By choosing a particular solution as
 \begin{equation}
 x_p^f={\beta^3t^2 \over 4m^2} + Ct,
 \end{equation}
and applying the unitary transformation to a stationary wave
packet $\psi^f (\equiv \exp[-{i\over \hbar}({\beta^3e \over 2m}t
+mCe)]\phi(x))$, one can find a wave packet $\psi_{free}$ for a
free particle system, as
\begin{eqnarray}
\psi_{free}&=&U^\dagger(x_p^f)\psi^f \cr
 &=&\exp[{i\over \hbar} \{ {m^3\over 3\beta^3}\{({\beta^3 t
 \over 2m^2}+C)^3 -C^3 \} -m({\beta^2t \over 2m^2}+C)(x+{\beta^3 t^2 \over 4m^2}
 +Ct+e)\}] \cr
 &&\times{\rm Ai}[-{\beta\over \hbar^{2/3}}(x+{\beta^3 t^2 \over 4m^2}
 +Ct+e)].
\end{eqnarray}
If we take $C=0$  and $e=0$, $\psi_{free}$ reduces to the wave
packet of ref. \cite{BB} which propagates in free space without
distortion and with constant acceleration. Our derivation of
$\psi_{free}$ shows that the solution for a free particle system
given in ref. \cite{BB} is related to the stationary wave packet
of zero energy-eigenvalue in a linear potential, while similar
wave packets for the free particle system can also be found from
the stationary wave packets of non-zero energy-eigenvlaues. By
using a unitary operator $U(x_p^f+D)$, one can obtain another
expression for $\psi_{free}$, which is, however, similar to the
one given in Eq. (12) with a redefinition of $e$. Even though the
Airy wave functions are square-integrable on a half-line ($x<L$)
\cite{CR} or on a piecewise domain, the wave function (packet) is
not square-integrable on the whole line, as have been discussed in
detail in ref. \cite{Landau}.

The wave function $\psi_M(\tau,x)$ satisfying the Schr\"{o}dinger
equation \[ i\hbar {\partial \over
\partial \tau}\psi_M(\tau,x)= -{\hbar^2\over 2M(\tau)}{\partial^2
\over
\partial x^2}\psi_M(\tau,x)\]
for a free particle with a time-varying mass $M(\tau)$, can be
found from $\psi_{free}$, by replacing $t$ with $\int_0^\tau
{m\over M(t)} dt$. Wave functions satisfying Eq. (1) can then be
found by applying the unitary operator to $\psi_M$, as
\begin{eqnarray}
 \psi(t,x)
 &=& U(x_p^M(t))\psi_M(t,x) \cr
 &=&\exp[{i\over \hbar}\{- {1 \over 2M(t')}
  \int_0^t(\int_0^{t'}F(t'')dt'')^2 dt'  +{1\over 3}({\beta^2 \over 2}\int_0^t {dt'\over
 M(t')}+{C\over \beta})^3   +\int_0^t xF(t')dt'  \cr
 & &~~~~~~~~~-({\beta^3\over 2}\int_0^t {dt'\over M(t')}+C)[x-x_p^M(t) +e
    +{\beta^3 \over 4}(\int_0^t {dt'\over M(t')})^2
             +C\int_0^t {dt'\over M(t')} ]\}]  \cr
 & &\times{\rm Ai}[-{\beta\over \hbar^{2/3}}(x-x_p^M(t)+e
       +{\beta^3 \over 4}(\int_0^t {dt'\over M(t')})^2
             +C\int_0^t {dt'\over M(t')})],
\end{eqnarray}
where
 \begin{equation}
  x_p^M(t)=\int_0^t{1\over M(t')} \int_0^{t'} F(t'') dt''dt'.
 \end{equation}
With the choice of $C=0$ and $e=0$, $\psi(t,x)$ reduces to the one
given in ref. \cite{Feng}.
\section{A complete set of wave functions}
The kernel of a harmonic oscillator described by the Lagrangian
 \begin{equation}
 L^F= {1\over 2} M(t) \dot{x}^2-{1\over 2}M(t) w^2(t)x^2 +F(t)x
 \end{equation}
whose classical equation of motion is given by
 \begin{equation}
 {d \over dt}(M{d\over dt}x) +w^2x=F,
 \end{equation}
has been given in ref. \cite{path}. The classical solution is
described by two linearly independent homogeneous solutions
$u_c(t)$ and $v_s(t)$ and one particular solution $x_{ph}(t)$. By
requiring the conditions $v_s(t_a)=0,~u_c(t_a)=1,~x_{ph}(t_a)=0,
~\dot{x}_{ph}(t_a)=0$ on classical solutions, the expression of
the kernel given in Eq. (32) of ref. \cite{path} can be simplified
to
 \begin{eqnarray}
 K^F(b,a)
 &=&\sqrt{{M(t_b)\over 2\pi i \hbar}{\dot{v}_s(t_a) \over v_s
 (t_b)}}\cr
 &&\times\exp[{i\over 2\hbar}\{x_a^2M(t_a)[-\dot{u}_c(t_a)
       + {u_c(t_b)\dot{v}_s(t_a) \over v_s(t_b)}]
       +[x_b-x_{ph}(t_b)]^2M(t_b){\dot{v}_s(t_b)\over v_s(t_b)} \cr
 &&~~~~~~~~~~~~~-2x_a(x_b-x_{ph}(t_b))M(t_a){\dot{v}_s(t_a) \over v_s(t_b)}
 +2M(t_b)\dot{x}_{ph} (t_b) x_b\cr
 &&~~~~~~~~~~~~~
 +\int_{t_a}^{t_b}(Mw^2x_{ph}^2-M\dot{x}_{ph}^2) dt
 \}],
 \end{eqnarray}
without loosing generality.
The kernel for the system described by the Hamiltonian $H$ can be
found by considering the harmonic oscillator of $w=0$. By letting
$u_c (t_b)=1$, $v_s(t_b)=\int_{t_a}^{t_b}{dt \over M(t)}$ and
$x_{ph}(t_b)= \int_{t_a}^{t_b}{1 \over
M(t)}\int_{t_a}^{t}F(t')dt'dt$, one may find that the kernel
(propagator) can be written as
 \begin{eqnarray}
 K(b,a)&=&K(t_b,x_b;x_a,t_a) \cr
 &=&\sqrt{{1\over 2\pi i\hbar}{1\over
 \int_{t_a}^{t_b}{dt \over M(t)}}} \cr
 &&\times
 \exp[{i\over 2\hbar}\{{1\over \int_{t_a}^{t_b}{dt \over
 M(t)}}(x_a-x_b
 +\int_{t_a}^{t_b}{1 \over M(t)}\int_{t_a}^{t}F(t')dt'dt)^2 \cr
 &&~~~~+2x_b\int_{t_a}^{t_b}F(t) dt
       -\int_{t_a}^{t_b}{1\over M(t)}(\int_{t_a}^{t}F(t')dt')^2dt
   \}].
 \end{eqnarray}
One can explicitly confirm that the kernel satisfies the initial
condition in the coincidence limit
 \begin{equation} \lim_{t_b \rightarrow t_a +0} K(b,a)=\delta(x_b-x_a),
 \end{equation}
and the Schr\"{o}dinger equation
 \begin{equation} i\hbar {\partial \over \partial t_b}K(b,a)
 = -{\hbar^2 \over 2M(t_b)}{\partial^2 \over
 \partial x_b^2}K(b,a) -x_b F(t_b) K(b,a).
 \end{equation}

The results for the harmonic oscillator system can also be used to
find wave functions for a particle in a linear potential. By
defining
 \begin{eqnarray}
 &&v(t)=\int_0^t{dt' \over M(t')} + p,   \\
 &&\rho(t)= \sqrt{v^2(t)+b^2},
 \end{eqnarray}
with a positive constant $b$ and an arbitrary real constant $p$,
the wave function $\psi_n$:
\begin{eqnarray}
 \psi_n(t,x)
 &=&{1\over \sqrt{2^nn!\sqrt{\pi\hbar}}}
    \sqrt{\sqrt{b}\over \rho}({b-iv \over \rho})^{n+{1\over2}}\cr
 &&\times
    \exp[{i\over \hbar}\{M\dot{x}_px+\xi(x_p)\}]
    \exp[{(x-x_p)^2\over 2\hbar}\{-{b\over
 \rho^2}+iM{\dot\rho \over \rho}\}]H_n(\sqrt{b\over
 \hbar}{x-x_p \over \rho})~~~~~~
\end{eqnarray}
may be given, where $H_n$ is the $n$-th order Hermite polynomial
and $x_p$ is given in Eq. (1). After some algebra, one can find
the relation
 \begin{equation} K(b,a)=\sum_{n=0}^\infty
 \psi_n(t_b,x_b)\psi_n^*(t_a,x_a),~~{\rm for}~~ t_b>t_a,
 \end{equation}
which proves that $\psi_n$ satisfies the Schr\"{o}dinger equation
of Eq. (1), and the set of $\{\psi_n | n=0,1,2,\cdots\}$ is
complete. While the relation in Eq. (24) is valid for a general
$M(t)$, by defining
 \begin{equation}
 z=\sqrt{b-i(t_b+p) \over b+i(t_b+p)}\sqrt{b+i(t_a+p) \over
 b-i(t_a+p)},
 \end{equation}
then Mehler's formula \cite{Erdelyi}
 \begin{equation}
 \sum_{n=0}^\infty{z^{n+{1\over 2}}\over 2^nn!}H_n(X)H_n(Y)
  =\sqrt{z \over 1-z^2}\exp[-{z^2\over 1-z^2}(X^2+Y^2)+2{z\over
 1-z^2} XY],
 \end{equation}
and the fact that
 \begin{equation}
{z\over 1-z^2}={\rho(t_b)\rho(t_a) \over 2ib(t_b-t_a)},
 \end{equation}
can be used for the proof of the relation for the case $M=1$.
Contrary to the case of the Airy wave function, the shape of the
probability distributions of these wave functions evolves as time
passes.
Following Ref. \cite{paraxial}, we define the generalized Gouy
phase factor $\chi(t)$ as
 \begin{equation} \tan\chi= {v\over b}.
 \end{equation}
Expressions for the generalized spot size $\gamma(t)$  and radius
of curvature of the wave front $s(t)$, which are real, are given
through a single complex equality
 \begin{equation}
 {1\over \gamma^2(t)}-{i\over \hbar s(t)}={1\over b+iv(t)}.
 \end{equation}
With these definitions, one can find that $\psi_n$ is written as
 \begin{eqnarray}
 \psi_n
 &=&{1\over \sqrt{2^nn!\sqrt\pi}}\cr
 & &\times{1\over \sqrt{\gamma}}
      \exp[{i\over \hbar}\{M\dot{x}_px+\xi(x_p)
      +{(x-x_p)^2 \over 2 s(t)}\} -i(n+{1\over 2})\chi -{(x-x_p)^2\over 2\gamma^2}]
 H_n({x-x_p \over \gamma}).~~~~
 \end{eqnarray}
For the case of $M=1$ and $F=0$, if we choose $x_p=0=p$, $\psi_n$
reduces to the Hermite-Gaussian mode in the paraxial approximation
\cite{paraxial}. On the other hand, it is clear that, by applying
the unitary transformation given in Eq. (6), the general
expression of $\psi_n(t)$ could be obtained from a Hermite-Gauss
mode whose time is $v$ $(=v(t))$.
\section{Summary} A unitary transformation which relates the
model of a linear potential (with time-dependent parameters) to a
free particle system has been given. This transformation closely
resembles the one responsible for the existence of coherent states
in harmonic oscillators, and the two arbitrary parameters in the
transformation have been interpreted as a manifestation of the
Galilean symmetry of classical mechanics. While this
transformation (with a change of the time-scale) can be used to
find a wave function of a particle in a linear potential from any
wave function of a free particle, this transformation has been
explicitly used to find a general wave packet described by an Airy
function. Based on the kernel and wave functions of a generalized
harmonic oscillator, the kernel and a set of Hermite-Gaussian type
wave functions are also given. The completeness of the set has
been proved for a general case, while such a proof is still not
available in the harmonic oscillator system.

\acknowledgments The author thanks Professor J.R. Klauder for a
critical reading and encouragement. This work was supported in
part by the Korea Research Foundation Grant (KRF-2002-013-D00025).

\end{document}